# A Stochastic Approach to Maxwell Velocity Distribution via Central Limit Theorem and Boltzmann's Entropy Formula

Sangita Mondal[‡] and Biman Bagchi[‡, *]

[‡]Solid State and Structural Chemistry Unit, Indian Institute of Science, Bengaluru, India

*Corresponding authors' Email: bbagchi@iisc.ac.in

## ABSTRACT

Maxwell's velocity distribution is known to be universally valid across systems and phases. Here we present a new and general derivation that uses the central limit theorem (CLT) of the probability theory. This essentially uses the idea that repeated intermolecular collisions introduce randomness in the velocity change in the individual components of the velocity vector, leading to, by the CLT, a Gaussian distribution. To complete the derivation, we next show that the mean-square velocity or the standard deviation follows exactly from Boltzmann's formula relating entropy to the density of states, thus avoiding the use of the ideal gas equation of state explicitly. We furthermore prove the Maxwell velocity distribution for a system of harmonic oscillators. This derivation provides a further insight into the origin of Boltzmann's constant in the Maxwell velocity distribution and also in the equipartition theorem. We propose that this derivation provides an approach that explains the universality of Maxwell gaussian distribution.

## I. Introduction

The velocity distribution of an interacting system at equilibrium is determined by its occupation probability density in the phase space. For a dilute gas, the situation is easy to comprehend. In this case, the system is mostly homogeneous, and the position variables are uniformly distributed. The velocity distribution is Gaussian, with the peak at the value of zero if we decompose into components. This is understandable because, in the absence of an external potential, there is no net motion of the system. Thus, the velocity distribution must be an even function such that positive and negative values are equally probable.

Here it is interesting to note that the Maxwell velocity distribution [1,2] is valid not only for gases, but also for liquids and solids, for all the classical systems. This fact plays an important role in diverse computer simulations where the equipartition theorem is used along with Maxwell velocity distribution to define the temperature. The validity of the Maxwell velocity distribution is really impressive if we analyze it a bit more. For example, the velocity time correlation function of a tagged particle differs considerably in going from gas to liquid to solid. In the gas phase, we can use Doob's theorem [3] that states that if the process is a gauss Markov process, then the time correlation of that quantity is exponential. The decay of velocity TCF is certainly not exponential in liquids and solids, although nearly so in the gas phase. However, the Maxwell distribution is found to hold exactly across all the phases, so long classical mechanics is followed.[4,5]

The standard derivations that are given in the undergraduate textbook on the kinetic theory of gases are not straightforward and use the ideal gas law at some stages.

Here we propose that a simpler derivation can be obtained by using the Central Limit Theorem of the probability theory. This is really straightforward.

An important step is the derivation of the mean square velocity. We show that this can be used by using the thermodynamic relation between entropy and energy, while the entropy can be obtained from the Boltzmann's formula[2] relating entropy with the density of states.

For dilute gas at equilibrium, with $N$ identical molecules occupying volume $V$ at temperature $T$, probability that the molecules have velocity in range $\mathbf{v}$ to $\mathbf{v}+\mathbf{dv}$ is given by [6,7]

$$f(\mathbf{v})d\mathbf{v} = 4\pi \left(\frac{m}{2\pi k_B T}\right)^{3/2} v^2 e^{-\frac{mv^2}{2k_B T}} d\mathbf{v} \tag{1}$$

The above expression is famously known as Maxwell-Boltzmann velocity distribution. We note that although students are introduced to this distribution as valid in the dilute gas, it actually is valid across classical systems and phases. It is accepted as universally valid, although the omnipresence of the Boltzmann constant $k_B$ could be a subject of curiosity and discussion.

In the three-dimensional velocity space, the components $v_x$, $v_y$, and $v_z$ of each molecule are distributed independently. Thus $f(\mathbf{v})$ can be written as the product of the probability densities of each component separately. Hence Probability that a molecule has velocity in the range $v_x$ to $v_x+dv_x$, $v_y$ to $v_y+dv_y$, $v_z$ to $v_z+dv_z$ is

$$f(v_x,v_y,v_z)dv_x dv_y dv_z = \left(\frac{m}{2\pi k_B T}\right)^{3/2} e^{-\frac{m[v_x^2+v_y^2+v_z^2]}{2k_B T}} dv_x dv_y dv_z \tag{2}$$

We now proceed to derive this distribution from the central limit theorem of probability theory. The proof has two distinct stages. First, we prove that the velocity distribution is Gaussian. Second, we prove that the standard deviation is indeed $\frac{k_B T}{m}$

## II. From the Central Limit Theorem

Central Limit Theorem states *that if a random variable X can be written as a sum of a large number of weakly correlated random numbers, then the distribution of X is gaussian*. Thus, if we can express X as [8]

$$X = x_1 + x_2 + x_3 + ...x_N \tag{3}$$

where $x$ $(i = 1, N)$ are uncorrelated or weakly correlated, then X is also a random variable that is described by a probability function $P(X)$ that is a Gaussian distribution. [9,10] That is, one can write the Probability distribution function of X as [11]

$$P(X) = \frac{1}{\sqrt{2\pi}\sigma} \exp^{-\frac{(x-\mu)^2}{2\sigma^2}} \tag{4}$$

Where σ is the standard deviation, equal to $\sqrt{\langle X^2 \rangle - \langle X \rangle^2}$ and $\mu = \langle X \rangle$.

## III. Maxwell velocity distribution from central limit theorem

In the case of gaseous and condensed phases, the velocity is continuously modulated by collisions, which are random, but must follow the momentum conversation laws. We can write the velocity at time t as follows

$$v_x(t) = v_x(t=0) + \delta v_1 + \delta v_2 .... + \delta v_N \tag{5}$$

*Since $\delta v_i$ are random and weakly correlated, then the probability distribution of $v_x$ at any time t is Gaussian.* The same holds for $v_y$ and $v_z$. *Note that the vector nature of velocity is of crucial importance in our proof.* A detailed discussion of the fluctuations is given in section V.

We next derive the expression for the standard deviation of this Gaussian velocity distribution.

## IV. Mean square velocity fluctuation from Boltzmann's entropy formula

What is still left is the derivation of mean square velocity fluctuation, that is, the standard deviation. Our derivation shall require two ingredients: (i) Boltzmann's formula relating entropy S to the density of microscopic states, and (ii) the thermodynamic relation relating entropy S, energy E, and temperature T.

We start from Boltzmann's entropy formula given by,

$$S = k_B \ln \Omega(N, V, E) \tag{6}$$

Where $\Omega(N, V, E)$ is the total number of microscopic states of the system kept at constant volume V, energy E with N as the total number of particles. We shall also need the thermodynamic expression that relates the entropy to temperature by using the well-known formula

$$\frac{1}{T} = \left(\frac{dS}{dE}\right)_{N,V} \tag{7}$$

### a. Non-interacting gas

We now proceed to evaluate the total number of states. In the absence of any spatial correlation among the particles, the total number of ways a particle can be distributed will be directly proportional to the volume of the phase space that provides a measure of the total number of microstates, $\Omega(N,V,E)$. Now for N atomic particles in three dimensions $\Omega(N,V,E)$ is given by [12]

$$\Omega(N,V,E) = \left(\frac{V}{h^3}\right)^N \frac{(2\pi mE)^{\frac{3N}{2}}}{\left(\frac{3N}{2}\right)!} \quad (8)$$

$$= \Gamma(V,N).E^{\frac{3N}{2}}$$

where, $\Gamma(V,N) = \left(\frac{V}{h^3}\right)^N \frac{(2\pi m)^{\frac{3N}{2}}}{\left(\frac{3N}{2}\right)!}$

The main point is that $\Gamma(V, N)$ contains no energy (E) dependence.

We next substitute the expression $\Omega(N,V,E)$ given in Eq. (8) to obtain

$$S = k_B \ln \Gamma + \frac{3}{2} N k_B \ln E \quad (9)$$

We take the derivative on both sides of Eq. (9) with respect to E to obtain

$$\left(\frac{dS}{dE}\right)_{N,V} = \frac{3}{2}\frac{Nk_B}{E} \quad (10)$$

We substitute the expression $\left(\frac{dS}{dE}\right)_{N,V}$ in Eq.(7) to obtain

$$\frac{1}{T} = \frac{3}{2}\frac{Nk_B}{E}, \text{ so } E = \frac{3Nk_BT}{2}. \quad (11)$$

Since the particles are non-interacting, thus only kinetic energy will contribute to the total energy. Therefore, E can be written as the sum of the individual components of kinetic energy.

$$E = \frac{3Nk_BT}{2} = N\frac{1}{2}m\left[\langle v_x^2\rangle + \langle v_y^2\rangle + \langle v_z^2\rangle\right] \quad (12)$$

The directions are equivalent, and the components of the mean squared velocity of the gas particles are equal and can be written as

$$\langle v_x^2\rangle = \frac{k_BT}{m} \quad (13)$$

*This completes our proof of the maxwell velocity distribution.* This also constitutes a proof of the equipartition theorem from Boltzmann's entropy formula.

### b. Solids: A system of harmonic oscillators

We now consider a system of non-interacting harmonic oscillators which can serve as a model of a solid phase via normal mode representation. Here we make the Einstein approximation that all the oscillators have the same frequency.[13] For a harmonic oscillator, the total number of states associated with energy E can be obtained by quadrature from the number of quantum states within a shell contained between the two constant-energy surfaces associated with energy $\varepsilon$ and $\varepsilon +d\varepsilon$. Let us consider a single harmonic oscillator of frequency $\omega_0$. The latter is given by the following well-known expression [14]

$$g(\varepsilon)d\varepsilon = \frac{\varepsilon^2 d\varepsilon}{2(\hbar\omega_0)^2}, \quad (14)$$

Thus, the total number of microstates for a single harmonic oscillator, $\Omega_1$, is given by

$$\Omega_1 = \int_0^E g(\varepsilon)d\varepsilon$$

$$= \int_0^E \frac{\varepsilon^2 d\varepsilon}{2(\hbar\omega_0)^2} \qquad (15)$$

$$= \frac{E^3}{6(\hbar\omega_0)^2}$$

$$= \Gamma'.E^3$$

Where, $\Gamma' = \dfrac{1}{6(\hbar\omega_0)^2}$.

Therefore, for N harmonic oscillators

$$\Omega_N = \Gamma'^N E^{3N} \qquad (16)$$

We now substitute the expression $\Omega_N$ given in Eq.(16) to obtain the following expression of entropy

$$S = Nk_B \ln\Gamma' + 3Nk_B \ln E \qquad (17)$$

We take the derivative on both sides of Eq. (17) with respect to E to obtain

$$\frac{1}{T} = \left(\frac{dS}{dE}\right)_{N,V} = \frac{3Nk_B}{E} \qquad (18)$$

$$E = 3Nk_B T \qquad (19)$$

Next, we note that the total E can be written as the sum of kinetic energy (K.E) and potential energy (P.E). For a harmonic oscillator, these two are equal. Therefore

$$2K.E = 3Nk_B T$$
$$N\left[2.\frac{3}{2}m\langle v_x^2\rangle\right] = 3Nk_B T \qquad (20)$$

Hence,

$$\langle v_x^2 \rangle = \frac{k_B T}{m} \qquad (21)$$

This is the expression for the standard deviation. And this then completes our proof of the Maxwell velocity distribution for a system of harmonic oscillators.

## V. Numerical Results and Discussion

We have performed standard molecular dynamic simulations [15] of 500 Lennard Jones particles for three different phases (gas, liquid and solid) at density-temperature combinations, (i) $\rho^* = 0.1$, $T^* = 2.5$, (gas) (ii) $\rho^* = 0.8$, $T^* = 0.4$, (iii) (liquid) $\rho^* = 0.8$, $T^* = 0.1$ (solid). We use the standard dimensionless (scaled) quantities, $\rho^* = \rho\sigma^3$ and $T^* = k_B T/\varepsilon$). The systems are equilibrated for two ns under NVT condition ($T^* = 2.5$, 0.4, and 0.1 for the three systems). Thereafter final production runs are carried out in the NVE ensemble. In **Figures 1**(a), (b), and (c), we show the velocity trajectories of a tagged particle for the three systems, respectively. The velocity fluctuates continuously and randomly because of interactions. The velocity fluctuations decrease at low temperature and high-density scenario. In **Figure 1**(d), (e), and (f), we depict the velocity distribution of the particle. The Gaussian nature of velocity distribution arises from the Central Limit Theorem. In **Figure 1**(g), (h), and (i), we show the normalized velocity autocorrelation function $C_v(t)$. Velocity autocorrelation functions are distinctly different for the three systems. It is observed that the velocity autocorrelation functions have backscattering for liquid the solid systems. In low temperature and high-density scenario, the collision time is very small, and surrounding liquid molecules form a cage. The negative region itself arises from the rebound of a molecule from the nearest neighbour molecules that form the cage. These have been discussed extensively in the past. We just present them here to substantiate our logic.

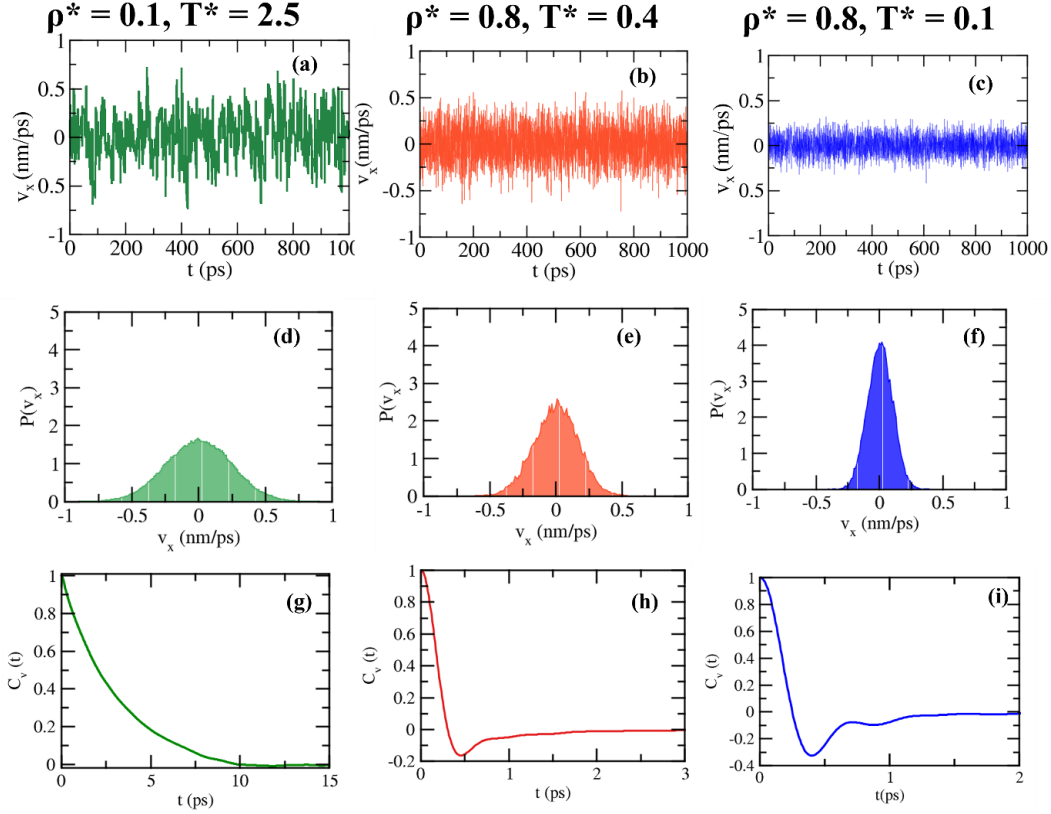

Figure 1: The molecular dynamic simulations are performed with 500 Lennard Jones particles for three different systems such as (i) $\rho^* = 0.1$, $T^* = 2.5$ (gas), (ii) $\rho^* = 0.8$, $T^* = 0.4$ (liquid), (iii) $\rho^* = 0.8$, $T^* = 0.1$ (solid). In Figure 1. (a), (b), and (c), we show the velocity values along trajectories of a tagged particle for the three systems, respectively. The velocity of the tagged particle fluctuates around a mean value. In Figure 1. (d), (e), and (f), we demonstrate the velocity distribution of them. The Gaussian nature of velocity distribution arises from the Central Limit Theory. The distributions become broad as we go from solid to liquid to gas. In Figure 1. (g), (h), and (i), normalized velocity autocorrelation functions are plotted. It is observed that the Velocity autocorrelation functions are distinctly different for the three systems.

## VI.    Summary

The robustness of the Maxwell velocity distribution is one of the central results of equilibrium statistical mechanics and of course, of the kinetic theory of gases. This is intimately connected with another central result which is the equipartition theorem. It is the generality of these two results, and their complete validity in all the phases of classical matter that motivated us to probe more deeply into the use of the central limit theorem of probability which is also a general and mathematically valid theorem. We note that the proof of the Maxwell distribution is well-established, through the perfect gas equation of state and the equipartition theorem. The present proof in terms of the Central Limit Theorem and Boltzmann's formula is exclusively from the first principle. It is entirely different and appears to be complete and robust. We have given proof both for ideal gas and also for a system of harmonic oscillators, the latter to model solids, and certainly more general. It captures the root cause of the universality of Maxwell velocity distribution in classical systems. It is clear that the use of the equipartition theorem provides an easy determination of the mean square fluctuation, however, often the equipartition theorem itself is presented through the Maxwell velocity distribution. Thus, we face here what can be called a tautology, this point deserves further study. Thus, our treatment here seems to provide an explanation of the equipartition theorem in the present context. The use of the central limit theorem to stochastic processes itself is certainly not new, but its use in providing an exact derivation of Maxwell velocity distribution, to the best of our knowledge, has not been reported. And also the use of Boltzmann's formula to obtain the standard deviation.

**ACKNOWLEDGEMENT**

We thank Mr. S. Acharya and Mr. S. Kumar for useful discussions. We are grateful to Professors R.N. Zare, D Wales, and S Saito for encouraging discussions and comments that have partly inspired this work. B.B. thanks SERB (DST), India, for the National Science Chair (NSC) Professorship and SERB (DST), India, for partial funding of this work. S.M. thanks IISc for the research fellowship.